\documentclass[useAMS,usenatbib]{mn2e}
\usepackage{epsfig,lscape} 
\usepackage{natbib}
\usepackage{amsmath}
\usepackage{comment}
\usepackage{hyperref}

\title[Origin of the $\alpha$-dichotomy and thick disc formation]{The age-chemical abundance structure of the Galactic disc II: $\alpha$-dichotomy and thick disc formation}

\author[J. Lian et al.]
{{Jianhui Lian}$^{1}$, {Daniel Thomas}$^{2}$, {Claudia Maraston}$^{2}$, Timothy C. Beers$^{3}$,
\newauthor {Christian Moni Bidin}$^{4}$, {Jos\'e G. Fern\'andez-Trincado}$^{5}$, 
{D. A. Garc\'ia-Hern\'andez$^{6,7}$}, 
\newauthor {Richard R. Lane}$^{5,8}$, 
{Ricardo R. Munoz}$^{9}$, 
{Christian Nitschelm}$^{10}$, 
{Alexandre Roman-Lopes}$^{11}$, 
\newauthor {Olga Zamora}$^{5,6}$
	\\
	\small $^{1}$Department of Physics \& Astronomy, University of Utah, Salt Lake City, UT, 84112, USA\\
	\small $^{2}$Institute of Cosmology and Gravitation, University of Portsmouth, Burnaby Road, Portsmouth, UK, PO1 3FX\\
	\small $^{3}$Department of Physics and JINA Center for the Evolution of the Elements, University of Notre Dame, Notre Dame, IN, 46556 USA\\
	\small$^{4}$Instituto de Astronomía, Universidad Católica del Norte, Av. Angamos 0610, Antofagasta, Chile\\
	\small $^{5}$Instituto de Astronom\'ia y Ciencias Planetarias, Universidad de Atacama, Copayapu 485, Copiap\'o, Chile\\
	\small $^{6}$Instituto de Astrofısica de Canarias, 38205 La Laguna, Tenerife, Spain \\
	\small $^{7}$Universidad de La Laguna (ULL), Departamento de Astrofísica, E-38206 La Laguna, Tenerife, Spain \\
	\small $^{8}$Instituto de Astrofısica, Pontificia Universidad Cat\'olica de Chile, Av. Vicuna Mackenna 4860, 782-0436 Macul, Santiago, Chile\\
	\small $^{9}$Departamento de Astronomia, Universidad de Chile, Camino El Observatorio 1515, Las Condes, Santiago, Chile\\
	\small $^{10}$Centro de Astronom{\'i}a (CITEVA), Universidad de Antofagasta, Avenida Angamos 601, Antofagasta 1270300, Chile\\
	\small $^{11}$Departamento de Física, Facultad de Ciencias, Universidad de La Serena, Cisternas 1200, La Serena, Chile\\
}

\begin{document}
\maketitle

\begin{abstract}	
We extend our previous work on the age-chemical abundance structure of the Galactic outer disc to the inner disc (4$<r<$8\;kpc) based on the SDSS/APOGEE survey. 
Different from the outer disc, the inner disc stars exhibit a clear bimodal distribution in the [Mg/Fe]--[Fe/H] plane. While a number of scenarios have been proposed in the literature, it remains challenging to recover this bimodal distribution with theoretical models. To this end, we present a chemical evolution model embedding a complex multi-phase inner disc formation scenario that matches {the observed} bimodal [Mg/Fe]--[Fe/H] distribution. 
In this scenario, the formation of the inner disc is dominated by two main starburst episodes $6\;$Gyr apart with secular, low-level star formation activity in between. In our model, the first starburst occurs at early cosmic times ($t\sim 1\;$Gyr) and the second one $6\;$Gyr later at a cosmic time of $t\sim 7\;$Gyr. Both these starburst episodes are {associated with} gas accretion events in our model, and are quenched rapidly. {The} first starburst leads to the formation of the high-$\alpha$ sequence, and the second starburst leads to the formation of the {\it metal-poor} low-$\alpha$ sequence. The {\it metal-rich} low-$\alpha$ stars, instead, form during the secular evolution phase between the two bursts. 
Our model shows that the $\alpha$-dichotomy originates from the rapid suppression of star formation after the first starburst. The two starburst episodes are likely to be responsible for the formation of the geometric thick disc ($z>$1\;kpc), with the {\it old inner} thick disc and the {\it young outer} thick disc forming during the first and the second starbursts, respectively.
%
%


\end{abstract}


\section{Introduction}
The Milky Way Galaxy (MW) is the ideal laboratory to study galaxy formation and evolution in great detail. Our ability to resolve individual stars allows us to unfold the Galactic evolutionary history with unprecedented accuracy in constraining key processes such as gas accretion, star formation, and chemical enrichment. 

The first census of stars in the MW was conducted for the solar  neighbourhood. Star counts in the vertical direction revealed the presence of two disc components with different scale heights. These were interpreted as geometric `thick' and `thin' discs \citep{yoshii1982,gilmore1983}. The presence of thick and thin disc components in other galaxies was first discovered by \citet{burstein1979}, and then confirmed to be ubiquitous in the local universe \citep{yoachim2006,comeron2011}. To explain the formation of the thick disc, many possible mechanisms have been proposed: 1) in-situ formation due to gas-rich mergers \citep{brook2004} or disc turbulence at high redshift \citep{noguchi1998,bournaud2009,clarke2019}, 2) vertical heating of thin disc stars \citep{quinn1993,villalobos2008}, 3) accretion from external satellites \citep{abadi2003}, or 4) radial migration of kinematically hot stars (\citealt{schonrich2009,loebman2011,roskar2013}, challenged however by \citealt{minchev2012}).

Interestingly, studies of the chemical compositions of stars in the solar  neighbourhood also identified the presence of two populations with {different} [$\alpha$/Fe] ratios \citep{fuhrmann1998,prochaska2000,reddy2006,lee2011,adibekyan2012,haywood2013,bensby2014}. This chemical division between disc stars is not identical, but shares significant overlap with the geometrical definition based on {vertical} height. Most importantly, the geometric thick disc component of the inner disc is dominated by the high-$\alpha$ sequence \citep{bovy2012b,hayden2015,martig2016b}. This implies that the mechanisms responsible for the thick disc formation and the $\alpha$-dichotomy are closely related.   



A number of recent, large spectroscopic surveys such as APOGEE \citep{majewski2017}, LAMOST \citep{zhao2012}, GALAH \citep{desilva2015}, and GAIA-ESO \citep{gilmore2012}, have expanded our horizon well beyond the solar  neighbourhood. Covering our Galaxy from the Galactic bulge in the centre to the edge of the disc at large radii, the $\alpha$-dichotomy is confirmed across the entire Galactic disc \citep{anders2014,nidever2014,hayden2015}. 
Such wide spatial coverage allows us to probe the geometric structure of each chemical population, providing further constraints on the origin of the $\alpha$-bimodality and the formation of the thick disc. \citet{bovy2012a} and \citet{bovy2012b} find that the vertical scale height of the high-$\alpha$ population does not change significantly with radius. This disfavours a radial migration origin of the thick disc. The same studies also find that scale height decreases with decreasing [$\alpha$/Fe] without a clear discontinuity. This suggests there might be no bimodality in the spatial structure. 

The origin of the $\alpha$-dichotomy has been an open question since its first discovery. 
A promising explanation is a two-phase star formation history. An early rapid star formation episode forms the high-$\alpha$ population, while a subsequent secular, long-lived phase produces the low-$\alpha$ population. \citet{chiappini1997} first proposed a `two-infall' scenario in a chemical evolution model, which considers two star formation episodes triggered by two gas infall phases with different timescales. This model qualitatively produces the general evolution in [$\alpha$/Fe]-[Fe/H] space \citep{anders2017}, but cannot reproduce the observed $\alpha$-dichotomy. Predictions from a revised `two-infall' model by \citet{spitoni2019} provide a better match to the observed age distribution but still does not match the observed double-peak in [$\alpha$/Fe]. In cosmological simulations, chemical bimodality is not usually produced \citep{loebman2011,minchev2013},
or, if present, it does not match the chemical abundances {and/or the shape of the bimodal distribution observed} in the MW \citep{grand2017,mackereth2018,clarke2019,vincenzo2020}. The observed $\alpha$-dichotomy remains a major challenge when simulating the chemical enrichment history of our Galaxy. 

By leveraging observations from asteroseismology observations, stellar ages are now measurable for a large sample of stars in modern spectroscopic surveys \citep{ness2016,martig2016a,wu2018,wu2019}. Combining the age and chemical abundance information allows us to directly unfold the chemical evolution history of our Galaxy.  
In a companion paper (\citealt{lian2020}, hereafter Paper I), we study the age-chemical abundance structure of the Galactic {\em outer} disc. An interesting coexistence of old metal-rich and young metal-poor populations is found, which we interpret as evidence of recent gas accretion likely induced by the accretion of a gas-rich dwarf galaxy, possibly the Sagittarius (Sgr) dwarf galaxy. 

In the present paper, we focus on the Galactic {\rm inner} disc, which shows a more complicated age-chemical abundance structure than the outer disc analysed in \citet{lian2020}. The aim is to unveil the underlying mechanisms responsible for the $\alpha$-dichotomy and the formation of the thick disc component. The structure of the paper is as follows: We briefly introduce the sample selection in \textsection2 and the main observational results in \textsection3. In \textsection4, we compare the predictions from our chemical evolution model with observations. We discuss the implications of our model on the origin of the $\alpha$-dichotomy and thick disc formation in \textsection5. Finally, a brief summary is provided in \textsection6.

\section{Data} 
The sample of MW stars is selected from the latest Sloan Digital Sky Survey (SDSS) data release 16 (DR16, \citealt{ahumada2019}), which contains 473,307 unique stars observed by the Apache Point Observatory Galactic Evolution Experiment (APOGEE) survey \citep{majewski2017}, {a core project of the SDSS-IV survey \citep{blanton2017}}. 
APOGEE targets primarily horizontal branch and red giant branch stars throughout the Milky Way's bulge, disc, and halo \citep{zasowski2013,zasowski2017}, using the 2.5~m Sloan Telescope and {the NMSU 1m Telescope} at the Apache Point Observatory \citep{gunn2006,holtzman2010}, the 2.5~m Ir\'en\'ee du~Pont telescope \citep{bowen1973} at Las Campanas Observatory, and a pair of high-resolution, $H$-band spectrographs \citep{wilson2012,wilson2019}. Data are reduced, and heliocentric velocities, stellar parameters, and chemical abundances are determined using the pipelines described in \citet{nidever2015} and \citet{garciaperez2016}. 


To ensure the high quality of the stellar parameter measurements, we apply a cut in signal-to-noise ratio (SNR): we select stars with APOGEE spectra of SNR above 60, and exclude stars with any potential caveats in the stellar parameter determination as flagged in the APOGEE DR16 catalogue. 
{Since  elemental  abundance  determinations  in  APOGEE  tend  to  be  less  reliable  at  low effective temperature, we further exclude stars with $T_{\rm eff}<3200\;$K. Our final sample consists of {39,548 stars.}}

\begin{figure}
	\centering
	\includegraphics[width=9cm]{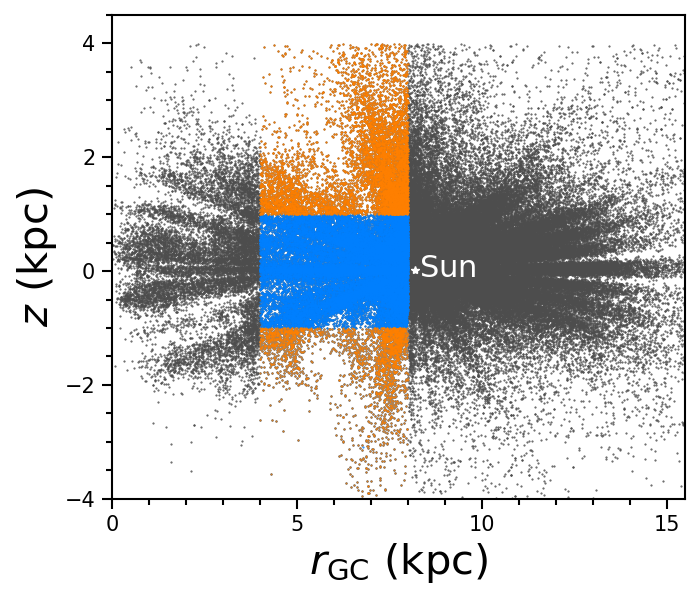}
	\caption{Spatial distribution of stars in the APOGEE DR16 catalog (grey) and our sample (orange and blue) in the $r$-$z$ plane. A vertical distance cut at 1 kpc is used to identify the geometric thin disc (blue) and the geometric thick disc (orange).}
	\label{spatial-dtr}
\end{figure}

We adopt the stellar parameters and abundances from the APOGEE DR16 catalogue, and spectro-photometric distances {with} the procedure described in \citet{rojas2017}. 
Since the magnesium abundance has been shown to be the most reliably measured $\alpha$-element in the APOGEE survey (\citealt{rojas2019}; J\"onsson et al. in prep), we use [Mg/Fe] as a tracer of [$\alpha$/Fe].
Stellar ages are taken from SDSS DR16 Value Added Catalog \citep{mackereth2019}\footnote{https://data.sdss.org/sas/dr16/apogee/vac/apogee-astronn}, which are derived with a Bayesian neural network model\citep{leung2019}\footnote{https://github.com/henrysky/astroNN} trained on asteroseismic ages \citep{pinsonneault2018}. 
{Note that the ages of stars become less reliable at [Fe/H]$<-0.5$, owing to the lack of training set stars and the effects of extra-mixing at these metallicities \citep{mackereth2019}.} The typical uncertainty in the derived age is 0.1 dex (i.e.,\ 25 per cent). 
Figure~\ref{spatial-dtr} shows the spatial distribution of all the stars in the DR16 catalog (grey points) in the $r$-$z$ plane. The blue and orange dots indicate the stars within $4<r<8\;$kpc analysed here selected to represent the inner disc. 

Based on the vertical density profile of Galactic disc stars {in the solar vicinity}, it has been shown that stars located at vertical distances larger than 1 kpc from the mid-plane are dominated by a second disc component, which is referred to as `thick disc'. The main disc component in the mid-plane is called the `thin disc' \citep{gilmore1983}. These are shown in orange and blue in Fig.~\ref{spatial-dtr}, respectively. 
Other commonly used definitions of the thin and thick discs are based on the dichotomy in chemical composition (e.g., \citealt{adibekyan2012}) or difference in kinematics \citep[e.g.,][]{robin2017}. 
In this paper, we adopt the geometric definition, and {refer to the `thick' and `thin' disc stars as stars above and on the mid-plane, respectively.}
We use the terms `low-$\alpha$' sequence and `high-$\alpha$' sequence, instead, to refer to the two populations of stars with {different} [$\alpha$/Fe]. 


\section{Chemical abundance patterns}
In the following, we identify and discuss qualitatively key chemical abundance patterns in the inner disc. These will be then be analysed {and interpreted} 
by means of a chemical evolution model in Section~\ref{sec:results}.

\subsection{[Mg/Fe]--[Fe/H]}
\begin{figure*}
	\centering
	\includegraphics[width=18cm]{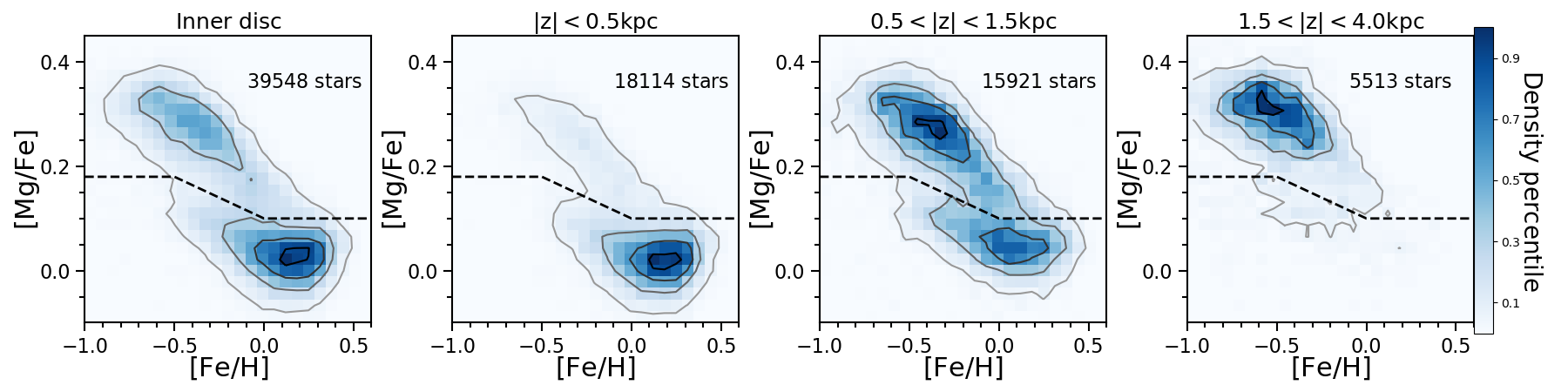}
	\caption{Distribution of inner disc ($4<r<8\;$kpc) stars in the [Mg/Fe]--[Fe/H] diagram. {\em Left-hand panel}: entire inner disc ($|z|<4\;$kpc), {\em centre-left panel}: {inner disc in the mid-plane} ($|z|<0.5\;$kpc), {\em centre-right panel}: {inner disc at intermediate height} ($0.5<|z|<1.5\;$kpc), {{\em right-hand panel}: inner disc above the mid-plane ($1.5<|z|<4\;$kpc).} The dashed line separates the low- and high-$\alpha$ populations. }
	\label{feh-afe-inner}
\end{figure*}

Figure~\ref{feh-afe-inner} shows the distribution of the Galactic inner disc stars in the [Mg/Fe]--[Fe/H] plane. The left-hand panel shows the entire inner disc, while the {three panels on the right-hand side are for stars at three different height bins, respectively. The number of stars in each height bin is indicated in each panel.}

The bimodal distribution in {the [Mg/Fe]-[Fe/H] plane} is evident (left-hand panel),{with a clear valley with low density between the two populations}. 
The separation is illustrated by the black dashed line, which is based on the separation line in \citet{adibekyan2011}. Its analytic form is:
\[
y = \left\{
\begin{array}{ll}
0.18,\ x<-0.5 \\
-0.16*x+0.1,\ -0.5<x<0 \\
0.1,\ x>0.
\end{array}
\right.
\]
In the following, we refer to stars above and below the dashed line as high-$\alpha$ and low-$ \alpha$ sequences, respectively.

The bimodal distribution of disc stars in [$\alpha$/Fe]-[Fe/H] was first identified in the solar neighborhood \citep{fuhrmann1998}, and then confirmed to be present across the inner disc based on large spectroscopic surveys \citep{hayden2015,wu2018}. It should be noted that this $\alpha$-dichotomy is most prominent in the inner disc, and gradually fades away toward large radii. This is because the density of the high-$\alpha$ population decreases faster with increasing radius $r$ than the low-$\alpha$ population. In other words, the radial scale-length of the high-$\alpha$ population is smaller \citep{bovy2012a,bovy2012b,bovy2016,mackereth2017}. In the outer disc, the bimodal distribution disappears completely, and the low-$\alpha$ stars become the dominant population \citep{hayden2015,lian2020}.

In addition to the variation in the radial direction, the distribution in [$\alpha$/Fe]-[Fe/H] also varies in the vertical direction along the $z$-axis. 
{The disc components in the mid-plane (centre-left panel) and above  the mid-plane (right-hand panel) are dominated by the low- and high-$\alpha$ populations, respectively. This suggests that the scale-height of the stellar component depends on [$\alpha$/Fe] (Bovy et al. 2012a, 2012b).}
Note, however, that the dependence of {scale-height on [$\alpha$/Fe] seems to be} a smooth function without a significant discontinuity {as seen in the [$\alpha$/Fe]-[Fe/H] plane} (Bovy et al. 2012a, 2012b).
 
\begin{figure*}
	\centering
你	\includegraphics[width=18cm]{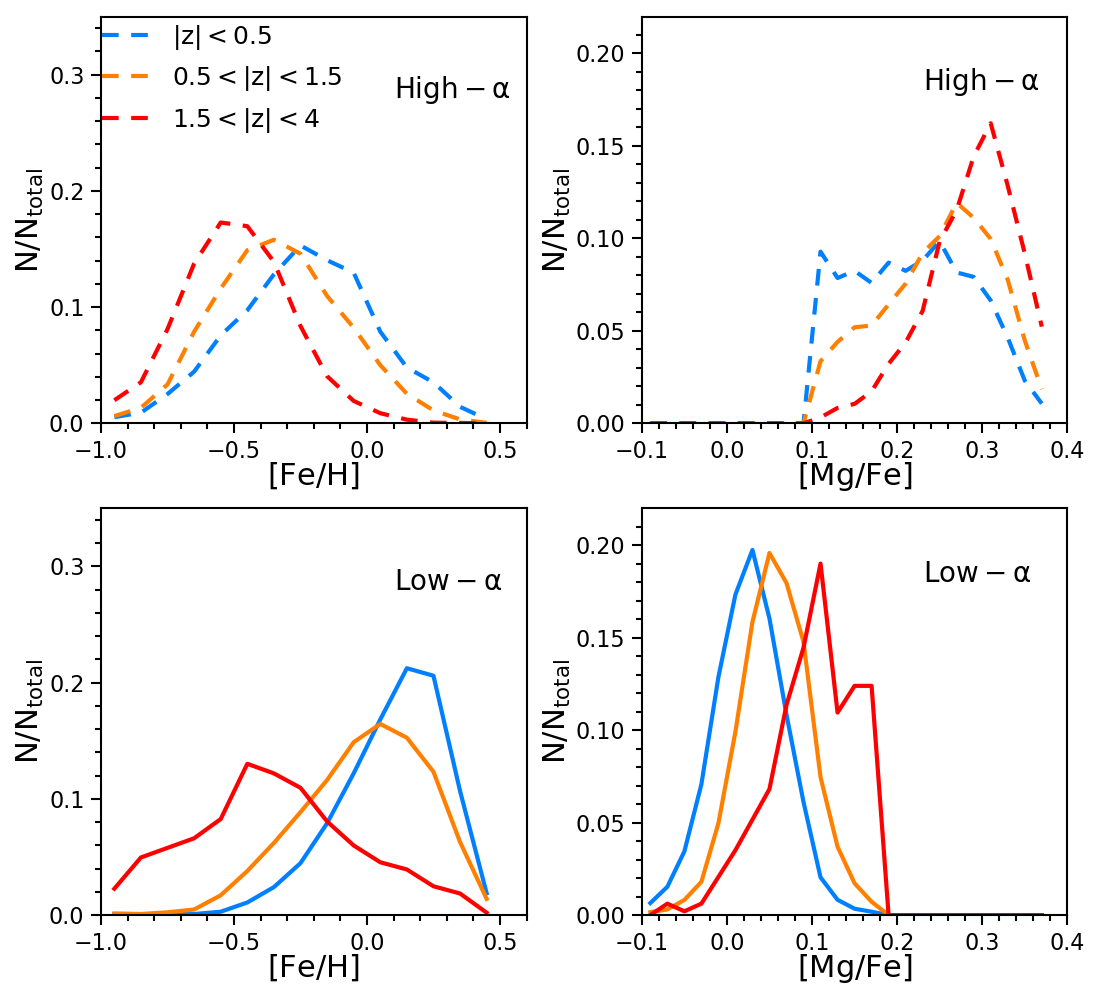}
	\caption{{Normalised distributions in [Fe/H] ({left-hand column}) and [Mg/Fe] ({right-hand column}) for the low-$\alpha$ sequence (bottom row) and the high-$\alpha$ sequence (top row) in the inner disc at different heights as indicated by the different colour. Both the low- and the high-$\alpha$ sequences show clear and consistent variations with height. Stars closer to the mid-plane tend to be more metal-rich and more $\alpha$-poor. 
	}}
	\label{hist-thick-thin}
\end{figure*}

\subsubsection{The high-$\alpha$ population}
The dependence of scale-height {on [Mg/Fe]} manifests itself in the shift of the low- and high-$\alpha$ sequences {observed at different heights}
in the [Mg/Fe]--[Fe/H] plane. The high-$\alpha$ sequence {in the mid-plane (centre-left panel)}
is slightly shifted toward higher [Fe/H] and lower [Mg/Fe] with respect to the high-$\alpha$ sequence {above the disc plane (centre-right and right-hand panels)}. 

To illustrate this more clearly, in Figure~\ref{hist-thick-thin} we plot the distributions in [Fe/H] and [Mg/Fe] for the high- and low-$\alpha$ populations {at different heights}. 
{The left-hand and right-hand columns show the distributions in [Fe/H] and [Mg/Fe], respectively. The distribution of the high-$\alpha$ sequence is shown in the top row, and the low-$\alpha$ population in the bottom row. Populations at different heights are indicated with different colour as illustrated in the legend. } 
Note that the sharp cut-offs in the [Mg/Fe] distribution of the high-$\alpha$ population {in the plane (blue dashed line in the top-right panel) and of the low-$\alpha$ population above the plane (red solid line in the bottom-right panel)}
are due to the separation of the high- and low-$\alpha$ populations based on [Mg/Fe]. 

The median [Mg/Fe] of the high-$\alpha$ population is {systematically} larger 
{at larger heights (0.22 for $z<0.5$, 0.26 for $0.5<z<1.5$, and 0.30 for $1.5<z<4\;$kpc).}
In addition, the median [Fe/H] {decreases with increasing height ($-0.23$ for $z<0.5$, $-0.35$ for $0.5<z<1.5$, and $-0.52$ for $1.5<z<4\;$kpc)).} 
These differences are statistically significant compared to their observational uncertainties, which are generally less than 0.02$\;$dex. This trend is consistent with the trend of decreasing scale-height with decreasing [$\alpha$/Fe] {and increasing [Fe/H]}, as found in Bovy et al. (2012b). An important implication of this trend is that the high-$\alpha$ sequence closer to the mid-plane is slightly more evolved (lower [Mg/Fe]), and could possibly have formed {following the formation of} the {more metal-poor} high-$\alpha$ sequence at large vertical distance. 
{Since high-$\alpha$} stars tend to be dynamically hotter with higher velocity dispersions and slower rotation speeds than the low-$\alpha$ stars \citep{bensby2004,adibekyan2012}, they are expected to {initially} form from highly turbulent gas 
{with wide vertical spread.} 
Without further energy injection, this metal-enriched gas will then dynamically cool and fall onto the disc mid-plane to fuel further star formation {therein}, leading to the formation of the {metal-rich} thin disc. This picture of two-phase star formation can naturally explain the more metal-rich and less $\alpha$-enhanced high-$\alpha$ sequence in the thin disc. 
We will explore this scenario 
with a chemical evolution model in Section~\ref{sec:results}. 

\subsubsection{The low-$\alpha$ population}
Like in the previous case, the low-$\alpha$ sequence is different {at different heights above the plane.} 
Fig.~\ref{feh-afe-inner} shows that {the low-$\alpha$ sequence tends to shift to lower [Fe/H] and higher [Mg/Fe] towards larger heights. This systematic shift is similar to that of the high-$\alpha$ sequence.}
Figure~\ref{hist-thick-thin} shows that the median [Mg/Fe] of the low-$\alpha$ population {increases with increasing height (0.03 for $z<0.5$, 0.06 for $0.5<z<1.5$, and 0.11 for $1.5<z<4\;$kpc)), while the median [Fe/H] decreases with increasing height (0.12 for $z<0.5$, 0.01 for $0.5<z<1.5$, and $-0.39$ for $1.5<z<4\;$kpc)).}
Again, this is in qualitative agreement with the trend of decreasing scale-height with decreasing [$\alpha$/Fe] {and increasing [Fe/H]} (Bovy et al. 2012a, 2012b).   

In Paper~I, we find a similar pattern for the outer disc at $10<r<15\;$kpc, where we identify the presence of metal-poor, low-$\alpha$ stars ([Fe/H]$<-0.2$, $[\alpha/{\rm Fe}]\sim +0.1$) {at $z>1\;$kpc}. 
Most surprisingly, these stars are younger than the more metal-rich stars. We propose that these metal-poor stars formed during a secondary starburst triggered by a recent major gas accretion event, likely triggered by the accretion of a gas-rich dwarf galaxy (see Paper~1 for details).
Interestingly, in the present paper we identify metal-poor, slightly $\alpha$-enhanced stars to be present at large vertical distances also in the inner disc. This suggests that the gas accretion event discussed in Paper~I may have also affected the inner disc, albeit less significantly.
 
%
%
       

\begin{figure*}
	\centering
	\includegraphics[width=18cm]{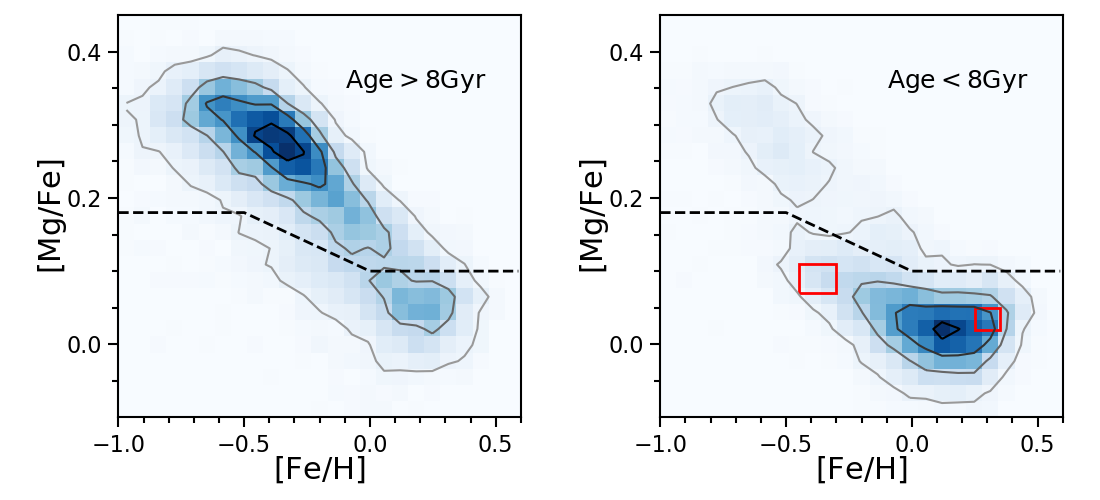}
	\caption{Distribution of inner disc stars in [Mg/Fe]--[Fe/H] split by age ({\em left-hand panel}: old stars with $t>8\;$Gyr, {\em right-hand panel}: young stars with $t<8\;$Gyr). The two {red} boxes in the right-hand panel indicate the low-$\alpha$ stars with the highest and the lowest [Fe/H], used to identify the populations formed at the beginning and end of a late-accretion episode (see text for more detail).}
	\label{afe-feh-agebin}
\end{figure*}

\subsection{Stellar ages}
\label{sec:ages}
In this section we add stellar age as a further parameter in our analysis of the [Mg/Fe]--[Fe/H] diagram. Figure~\ref{afe-feh-agebin} shows the distribution in [Mg/Fe]--[Fe/H] for {the relatively} old ($t>8\;$Gyr, left-hand panel) and young ($t<8\;$Gyr, right-hand panel) populations in the entire inner disc. The black dashed line is the same as Fig~\ref{feh-afe-inner} to separate the low- and the high-$\alpha$ sequences.

As expected, stars in the high-$\alpha$ sequence are generally older than stars in the low-$\alpha$ sequence. The bimodal distribution in [Mg/Fe]--[Fe/H] is replicated for both the old and the young populations, even though there are relatively few low-$\alpha$ stars in the old population (left-hand panel), and likewise relatively few high-$\alpha$ stars in the young population (right-hand panel). The presence of the latter is very interesting, though, as it challenges traditional chemical evolution models of the MW. We expect the majority of these stars in Figure~\ref{afe-feh-agebin} to be artefacts caused by the relatively large uncertainty in the age measurements of the present data set. 
Note, however, that a very small number of such stars have been identified with accurate seismic age measurements in the solar neighborhood \citep{chiappini2015,martig2015}. 

In the old population (left-hand panel), a notable fraction of stars is located in the low-$\alpha$ sequence with super-solar [Fe/H]. 
Such old, metal-rich stars are also present in the outer disc, as discussed in Paper I. 
However, they are predominantly found at small radii, hence are most prevalent in the inner disc \citep{bovy2016,mackereth2017}.
Interestingly, the density of stars does not drop to zero between the high- and low-$\alpha$ population, suggesting an evolutionary connection between them. Further evidence supporting this connection is their co-existence in the disc with similar scale-lengths \citep{bovy2016,mackereth2017}. 
In the next section, we will use our chemical evolution model to explore the potential evolutionary connection between the two populations, with the aim to shed light on the origin of the $\alpha$-dichotomy.  

As discussed above, the secondary accretion of gas included in our model of the outer disc very likely affected also the inner disc. As described in detail in Paper~I, the populations formed at the beginning and end of this late-accretion episode can be identified as the low-$\alpha$ stars with the highest and the lowest [Fe/H], respectively. This is because, in this scenario, metal-poor gas is being accreted leading to a depletion of [Fe/H]. Hence, the ages of these two populations indicated by the two {red} boxes in the right-hand panel of Figure~\ref{afe-feh-agebin} provide a rough constraint on the epoch of the late accretion event and its timescale. 
We estimate their average ages as $6.15\pm 1.90\;$Gyr and $5.85\pm2.04\;$Gyr, respectively, with an age difference of $0.30\pm 2.79\;$Gyr. Given the considerable scatter, the age difference is rather uncertain and not statistically significant. We will use this age difference as a reference instead of a strict constraint to the chemical evolution model. 

\section{Results}
\label{sec:results}
In this section we explore disc formation scenarios by means of a chemical evolution model constrained by the observed abundance patterns presented in the previous sections. 

\subsection{Chemical evolution model}
The chemical evolution model in this work was initially developed to investigate the mass-metallicity relation of galaxies \citep{lian2018a} and its cosmic evolution \citep{lian2018c}, as well as the radial distribution of gaseous and stellar metallicities within local galaxies \citep{lian2018b,lian2019}. In Paper I, we expanded this model to further consider the chemical enrichment histories of individual elements to aid the interpretation of observed chemical compositions of individual stars in our Galaxy. We briefly introduce the key ingredients of our model here, and refer the interested reader to \citet{lian2018a} and Paper I for more details.

The model considers three basic processes that regulate metal enrichment: gas accretion, star formation, and gas outflow. The star formation in the model follows the Kennicutt-Schmidt (KS) star formation law \citep{kennicutt1998}, which relates the SFR surface density with the gas mass surface density following
\begin{equation*}
\Sigma_{\rm SFR}=2.5\times10^{-4}\times C_{\rm ks}\times(\Sigma_{\rm gas}/{\rm M_{\odot}pc}^{-2})^{n_{\rm ks}} {\rm M_{\odot}yr^{-1} kpc}^{-2}.
\end{equation*}
We fix the power index, $n_{\rm ks}$, to be 1.5. The coefficient $C_{\rm ks}$, which regulates the star formation efficiency (SFE), is normalised to the original value in the KS law (i.e.,\ $2.5\times10^{-4}$), and is a free parameter in the model. We adopt the Kroupa initial mass function (IMF; \citealt{kroupa2001}). Metal production from asymptotic giant branch (AGB) stars, Type-Ia supernovae (SN-Ia), and Type-II supernovae (SN-II) is considered in the model. The yields table used for each process are explained in \citet{lian2018a}, except that the SN-II yields have been updated following \citet{kobayashi2006} (hereafter K06). 
We note that the magnesium abundance in the K06 yields is underestimated relative to oxygen by $\sim 0.1$ dex (see discussion in \textsection5.1). We therefore increase the magnesium yields by this amount. 

To account for the SN-Ia rate, a direct approach is to model the SN-Ia rate based on a theoretical SN-Ia rate formalism, with assumptions on the progenitor type of SN-Ia, the binary mass function, the secondary mass fraction distribution, and binary lifetimes \citep{greggio1983,matteucci1986,thomas1998}. Due to the uncertain nature of SN-Ia progenitors, many empirical SN-Ia delay-time distributions (DTDs) have been proposed. These are calibrated with the observed SN-Ia rates \citep{strolger2004,matteucci2006,maoz2012}. In the present work, we adopt the power-law SN-Ia DTD by \citet{maoz2012}:
\begin{equation*}
{\rm DTD} = a(\tau/{\rm Gyr})^{-1.1}, 
\end{equation*}  
where $\tau$ is the delay time of the SN-Ia explosions, and $a$ is a normalisation constant.  

By modelling gas and stellar metallicities of local star-forming galaxies in \citet{lian2018a,lian2018b} we found that the strength of metal outflow (i.e.,\ the metal mass loading factor) {anti-correlates with} a galaxy's total stellar mass, {probably due to mass-dependent gravitational well depth}, and plays a relatively unimportant role in the chemical evolution of massive star forming galaxies with stellar mass above $10^{10.5}\;{M_{\odot}}$. {This mass-dependent outflow driven by supernovae is also confirmed in cosmological simulations \citep{nelson2019}.} Assuming a stellar mass of $\sim 6\times 10^{10} M_{\odot}$ for our Galaxy \citep{mcmillan2011}, we therefore do not include any metal outflow in our model. {Note that the metal outflow could be important for the lower-mass, high-$z$ progenitors of today's massive star-forming galaxies. This would only affect the early enrichment history of our Galaxy with no significant impact on our results.}


{Stellar radial migration (changing guiding radius) is one of the radial mixing processes of stars commonly seen in dynamical simulations and is thought to play an important role in re-distributing stars in the Milky Way \citep{schonrich2009,minchev2013}. However, the overall impact of radial migration across the Galactic disc is still controversial. \citet{haywood2013} and \citet{snaith2014} find that the chemical structure within and beyond the solar vicinity could be explained by a multi-phase formation history without the need for radial migration (see opposite views in \citet{minchev2014}, \citet{frankel2019} and \citet{sharma2020}). To study the potential effect of radial migration on the age-chemical abundance structure of the Galactic disc, we explored a simple model that accounts for radial migration in Paper I. 
By comparing to the observations in age-[α/Fe]-[Fe/H] space, we found that the complex age-chemical abundance structure of outer disc can not be explained by the radial migration alone but serves as strong evidence for a late accretion/burst event. In this work we aim to propose a viable rather than an exclusive scenario that can explain the abundance pattern observed in the inner disc. In light of Paper I and previous analyses on the chemical abundance structure of the disc \citep{haywood2013,snaith2014}, we therefore do not consider radial migration in the present analysis. 
}

\subsection{Viable scenario}
\begin{figure*}
	\centering
	\includegraphics[width=20cm,viewport=130 10 1200 260,clip]{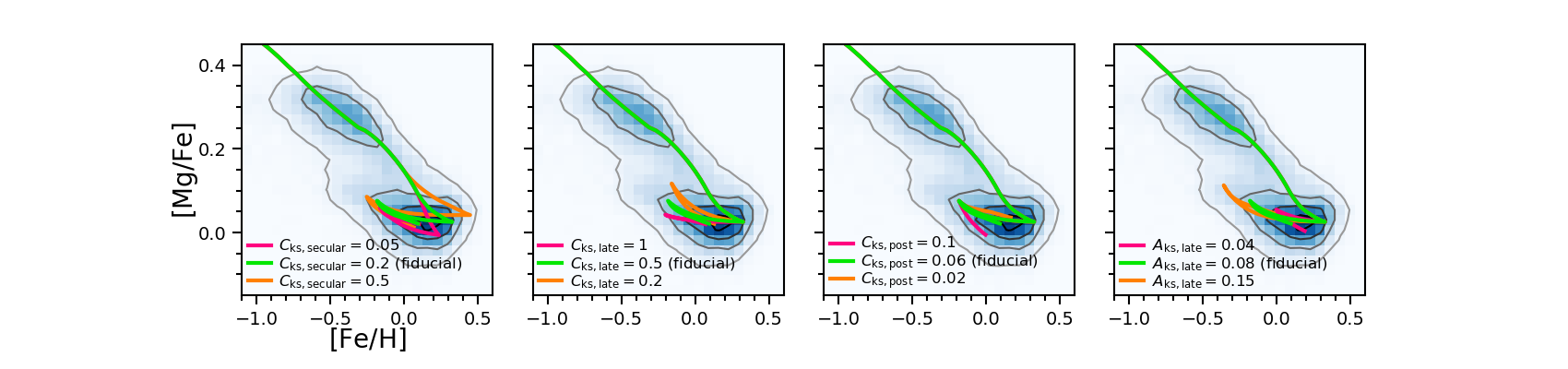}
	\caption{Effect of varying model parameters that characterise gas accretion and star formation during the secular and late-accretion phases 
	on the model track in the [Mg/Fe]-[Fe/H] plane. Each panel shows models varying one parameter as indicated in the legend. The underlying density map and black contours are the observations.}
	\label{feh-mgfe-models-dr16}
\end{figure*}

{We explored the parameter space of multiple-phase gas accretion scenarios and associated star formation histories, in order to find a coherent model for the formation of the MW disc. The main parameters are the timescale and the strength of each phase of gas accretion and star formation. Inspired by Paper I, we find a scenario that consists of two phases of gas accretion and associated star bursts that are 
separated by a prolonged period of low-level star formation, which is able to reproduce the observed pattern in the [$\alpha$/Fe]-[Fe/H] plane, including the $\alpha$-dichotomy.

In this scenario, the high- and low-$\alpha$ sequences form during these two separate starbursts occurring at different epochs.  
We refer to this scenario as the `two-burst' scenario. The gas accretion and star formation histories of this two-burst scenario are characterised by nine parameters.
\begin{enumerate}
    \item The gas accretion rate in the initial accretion event ($A_{\rm initial}$)
    \item The gas accretion rate in the second accretion event ($A_{\rm late}$)
    \item The start time of the late-time accretion event ($t_{\rm late}$)
    \item The length of the first accretion event ($L_{\rm initial}$)
    \item The length of the second accretion event ($L_{\rm late}$)
    \item The normalised coefficient of the star formation law during the initial burst ($C_{\rm ks,intial}$)
    \item The subsequent secular phase ($C_{\rm ks,secular}$)
    \item The late-time burst ($C_{\rm ks,late}$)
    \item The subsequent post-burst phase ($C_{\rm ks, post}$).
\end{enumerate}
    
There is a mild degeneracy between the parameters describing the first accretion/burst event ($A_{\rm initial}$ and $C_{\rm ks,intial}$) in reproducing the observed trend of the high-$\alpha$ sequence. A higher initial accretion rate would lead to a higher initial star formation efficiency that will predict a too high [Mg/Fe] at a given [Fe/H]. This effect could be balanced by a lower $C_{\rm ks,intial}$. 
The length of the initial burst ($L_{\rm initial}$) is relatively well constrained due to the rapid enrichment speed at this stage. 
A difference of 0.2$\;$Gyr in $L_{\rm initial}$ will result in a shift of the peak density of high-$\alpha$ sequence by $\sim0.1\;$dex in [Fe/H]. Overall the initial accretion/burst event is not strictly constrained by the data presented here. Observations at lower metallicity that cover the [$\alpha$/Fe] plateau would provide stronger constraints on these parameters.  

In contrast to the initial one, the late-time accretion/burst event is relatively well constrained by the data. 
The onset and end of the late-time accretion and burst event ($t_{\rm late}$ and $L_{\rm late}$) are largely constrained by the ages of the stars in the low-$\alpha$ sequence as explained in Paper I and Fig.~\ref{afe-feh-agebin}. The uncertainties of the absolute timing and duration of this accretion/burst event are therefore comparable to uncertainty of the age. The remaining four parameters are constrained by the observed `banana' shape of the low-$\alpha$ sequence. To demonstrate how these parameters are constrained and how models will change when varying these parameters, we present four sets of models in Figure~\ref{feh-mgfe-models-dr16} with each set shown in each panel varying one parameter. These models also illustrate the allowed ranges of these parameters. The underlying density map and contours show the observations of the inner disc which are the same as the left-hand panel of Fig.~\ref{feh-afe-inner}.

Compared to the reference models (magenta and orange lines) in Fig.~\ref{feh-mgfe-models-dr16}, the fiducial model (green line) best matches the overall shape of the low-$\alpha$ sequence. 
As will be discussed in \textsection4.3.2, it also provides a good match to the observed distribution functions in [Mg/Fe] and [Fe/H]. 
Since the goal of this paper is to present a viable scenario that can explain the chemical abundance pattern observed in the inner disc, the fiducial model is chosen based on visual inspection on the goodness of match between model and data. The parameters adopted for the fiducial model are listed in Table~\ref{table1}.

This two-burst scenario scenario is different from the traditional `two-infall' model, in that it implies the first star formation epoch to be quenched early, followed by a second late-accretion event several Gyr later. 
These two starburst phases are bridged by a prolonged phase of low-level star formation. In the traditional two-infall model, instead, the second accretion occcurs at earlier epochs, and on longer timescales \citep{chiappini1997,spitoni2019}.

It is worth pointing out that the low-$\alpha$ sequence shifts systematically towards lower metallicity with increasing radius \citep[e.g.,][]{hayden2015}. Based on the two-burst scenario presented here, this radial trend in the low-$\alpha$ sequence can be explained if the disc is not homogeneously affected by the late-accretion event. A possible solution is the accretion of more metal-poor gas onto the disc at larger radii, as shown in the right-hand panel of Fig.~\ref{feh-mgfe-models-dr16}.   

\begin{table*}
	\caption{Model parameters adopted for the fiducial model in this work. Parameter definitions are described in \textsection4.2.
	}
	\label{table1}
	\centering
	\begin{tabular}{l c c c c c c c c c}
	\hline\hline
	& $A_{\rm initial}$ & $C_{\rm ks,intial}$ & $L_{\rm initial}$ & $C_{\rm ks,secular}$ & $A_{\rm late}$ & $C_{\rm ks,late}$ & $t_{\rm late}$ & $L_{\rm late}$ & $C_{\rm ks, post}$ \\
	& ${\rm M_{\odot}yr^{-1}kpc^{-2}}$ & - & Gyr & - & ${\rm M_{\odot}yr^{-1}kpc^{-2}}$ & - & Gyr & Gyr & -  \\
	\hline
	Fiducial model & 0.05 & 1.5 & 1.3 & 0.2 & 0.08 & 0.5 & 7 & 0.5 & 0.02 \\
	\hline
	\end{tabular}\\  
\end{table*}
}

\subsection{Model predictions}
\begin{figure*}
	\centering
	\includegraphics[width=18cm]{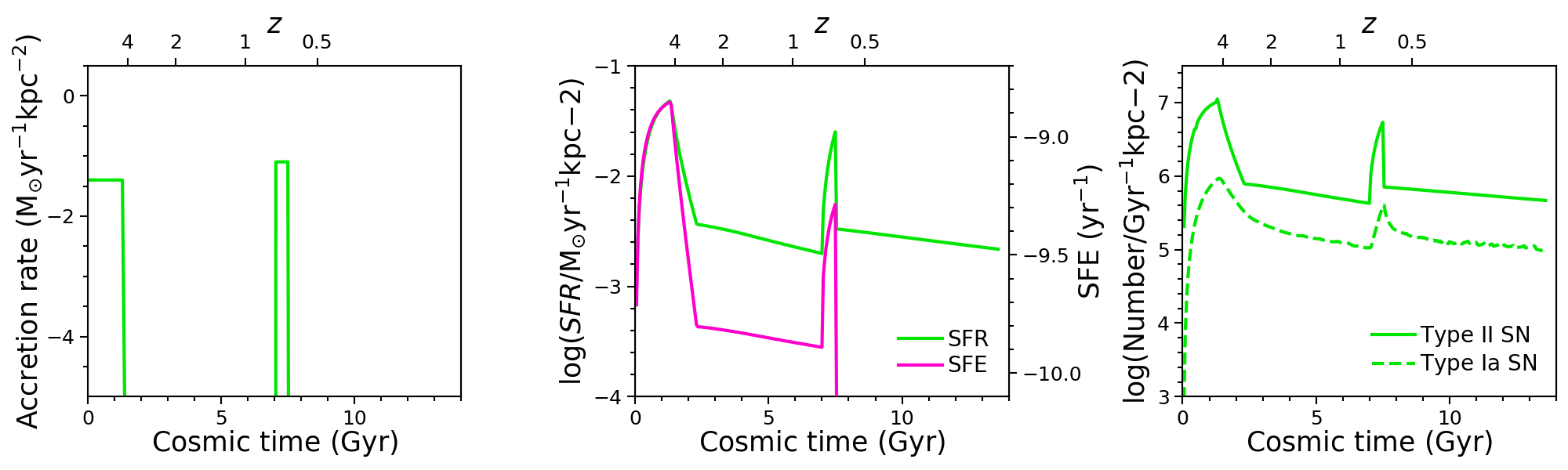}
	\caption{Gas accretion rate (left-hand panel), star formation rate SFR (middle panel, green line), star formation efficiency SFE (middle panel, magenta line), and supernova rates (right-hand panel) as a function of cosmic time, as predicted by our {fiducial} model. }
	\label{sfh}
\end{figure*}


In the following, we present the prediction of our {fiducial} model for the time evolution of the gas accretion rate, the star formation rate, the star formation efficiency, and supernova rates, as well as the element abundance ratios [Mg/Fe] and [Fe/H].

\subsubsection{Gas accretion and star formation histories}
\begin{figure*}
	\centering
	\includegraphics[width=18cm]{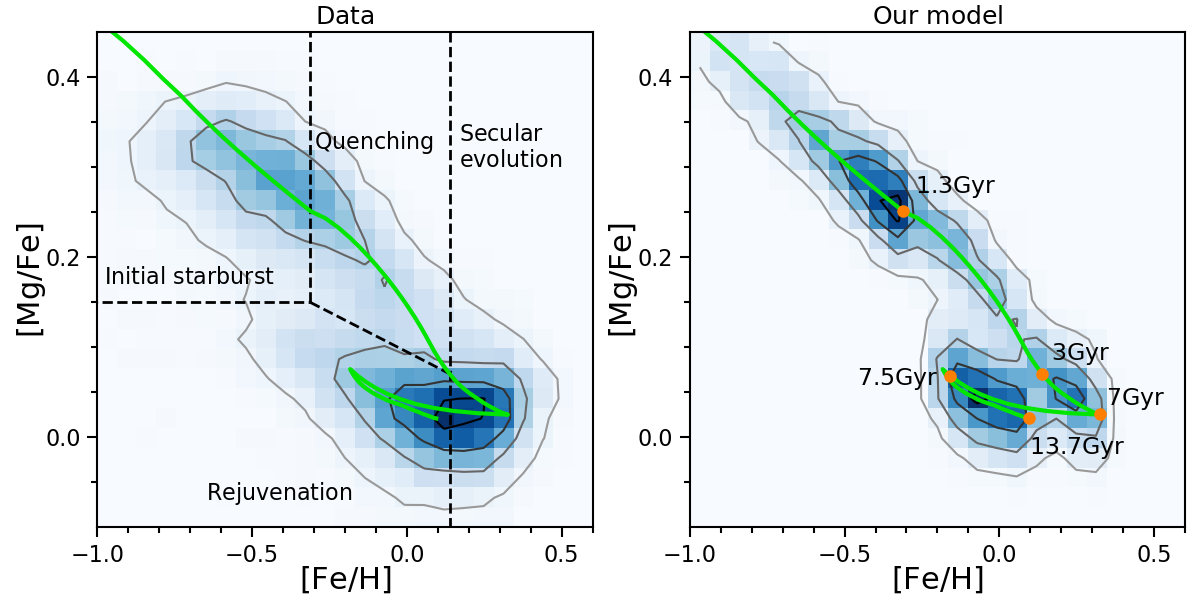}
	\caption{Density distribution of stars in the [Mg/Fe]--[Fe/H] plane as observed (left-hand panel) and predicted by our {fiducial} model (right-hand panel). The green line denotes the trajectory of the model. 
	Important time nodes in the model are marked as large orange circles along the track. Based on the simulation, the [Mg/Fe]--[Fe/H] plane can be divided into four sections: initial starburst, star formation quenching, secular evolution, and recent rejuvenation.}
	\label{feh-afe-model}
\end{figure*}

Figure~\ref{sfh} shows the predicted gas accretion history (left-hand panel), the SFR and SFE (middle panel), and the SN-II and SN-Ia rates (right-hand panel).  
An initial gas accretion phase with constant accretion rate is assumed for the first $1.3\;$Gyr (left-hand panel). After that, no gas is accreted until a second accretion event at a cosmic time of $7\;$Gyr. The timescale of this second accretion event is only $0.5\;$Gyr. 
Note, however, that the exact value of this timescale is quite uncertain, because of the considerable uncertainty in the age determination (see discussion in Paper~I). The timescale is constrained within about 1\;Gyr. Models with secondary accretion as short as 0.05\;Gyr or as long as 1.5\;Gyr still match the data. More accurate ages are needed in the future to better constrain the length of this second accretion event. 

During the first phase of gas accretion, both the SFR and SFE increase steeply, leading to the first starburst (middle panel). The latter is triggered by the high gas density resulting from the rapid accretion of gas. 
After reaching a peak at $1.3\;$Gyr, the SFR and the SFE drop quickly to level off at $\sim 2.5\;$Gyr. This first starburst phase is then followed by several Gyrs of low-level star-formation activity, with a slowly declining SFR and SFE. This rapid decline is essential to produce a density valley in the [$\alpha$/Fe]-[Fe/H] plane (see Section~\ref{sec:mgfe-feh}).

This phase of secondary evolution comes to an end with the onset of the second accretion even at $7\;$Gyr. A second star burst is triggered, and the SFE increases by a factor of $\sim 10$. This relatively recent star formation episode leads to the peculiar age-chemical abundance structure of the low-$\alpha$ sequence stars (see Section~\ref{sec:ages} and Paper~I). It is also consistent with the age distribution function of local white dwarfs (\citealt{vergely2002,cignoni2006,rowell2013,fantin2019}). A recent study of solar  neighbourhood star formation history using {\sl Gaia} observed color-magnitude diagram reveals a few recent starbursts that is possibly associated with Sgr pericentre passages \citep{ruiz-lara2020}. Among these starbursts, the strongest one occurred at 5.7$\;$Gyr ago which is well consistent with the second starburst identified from chemical abundance pattern in this work. Note that the second accretion and starburst of the model for the inner disc starts $\sim1\;$Gyr earlier than that in the model for the outer disc in Paper I. This is due to difference in ages used in this work and Paper I which are determined based on different methodologies and data.  


\subsubsection{Abundance ratios [Mg/Fe] and [Fe/H]}
\label{sec:mgfe-feh}

\begin{figure*}
	\centering
	\includegraphics[width=16cm]{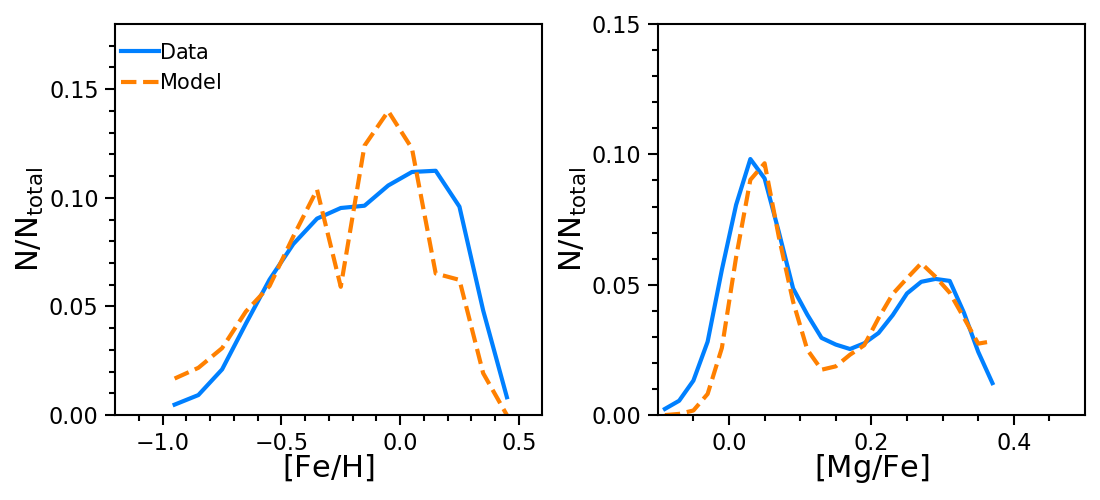}
	\caption{Comparison between the observed and predicted distributions in [Fe/H] (left-hand panel) and [Mg/Fe] (right-hand panel). The orange dashed lines are the model predictions, and the blue solid lines are the observations.}
	\label{hist-feh} 
\end{figure*}   

\begin{figure*}
	\centering
	\includegraphics[width=18cm]{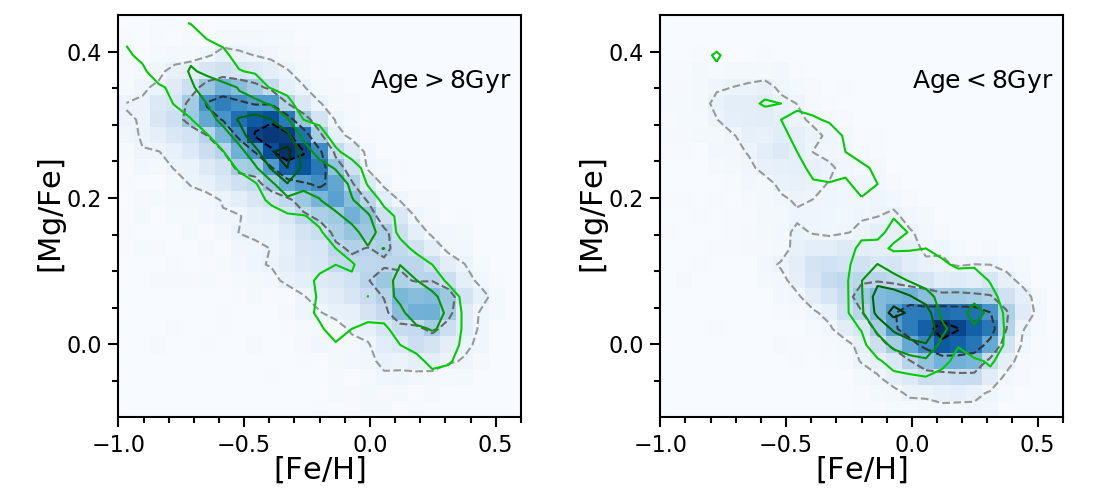}
	\caption{Comparison between the observed and predicted distributions in the [$\alpha$/Fe]-[Fe/H] plane for the old (left-hand panel) and young (right-hand panel) populations. Grey contours show the observed distribution which is the same as Fig.~\ref{afe-feh-agebin}. Green contours are the model predictions. }
	\label{agebin-model}
\end{figure*}

The [Mg/Fe]--[Fe/H] plane provides powerful constraints on the underlying chemical evolution model, and is widely used in the literature. The extensive and near-complete mapping of the MW with the new large-scale surveys now allow us to push this further, and to use also the density distribution of stars along the [Mg/Fe]--[Fe/H] relation. A key feature is the observed $\alpha$-bimodality that provides further key constraints on the star formation and the chemical enrichment histories.

Figure~\ref{feh-afe-model} shows the observed (left-hand panel) and predicted (right-hand panel) density distributions in the [Mg/Fe]--[Fe/H] plane. The trajectory of the {fiducial} `two-burst' model is shown by the green line. Observational uncertainties have been considered in the density distribution of the simulation. We adopt an uncertainty of 0.02 dex for both [Mg/Fe] and [Fe/H], and an uncertainty of 0.1 dex for age. 
Key epochs are marked as large orange circles along the model trajectory. 


A clear bimodal distribution is present in the model, matching the observed distribution well. The high- and low-$\alpha$ sequences in the model originate from the two starbursts occurring at 1.3 and 7 Gyr, respectively. The shift in [Mg/Fe] between the two sequences is a consequence of the rapid suppression of star formation after the first starburst.

Another effect of the rapidly decreasing SFR is the low number of stars formed, which is critical to produce the dearth of stars with intermediate [Mg/Fe]. Residual star formation after the rapid decreasing phase lasts $\sim 4$\;Gyr, which leads to the formation of the metal-rich part of the low-$\alpha$ sequence (see discussion in \textsection3.4). 

The low-$\alpha$ sequence forms in the second gas accretion event at $\sim$7\;Gyr.
%
During this episode, the accretion of a large amount of pristine gas dilutes the overall metallicity in the gas, and leads to a decrease in iron abundance. Therefore, the model track evolves backward to lower [Fe/H]. The enhanced star formation triggered by the gas accretion in turns leads to a slight increase in [Mg/Fe]. Therefore, the amount of gas accreted during the second accretion episode is directly constrained by the low-metallicity end of the low-$\alpha$ sequence, while the enhancement in SFE is constrained by the corresponding (maximum) [Mg/Fe] of the low-$\alpha$ sequence.

Following the star formation history shown in Figure~\ref{sfh}, we split the [Mg/Fe]--[Fe/H] diagram into the following four sections:
\begin{enumerate}
    \item initial starburst,
    \item rapid star formation quenching,
    \item long-lived secular evolution,
    \item recent rejuvenation.
\end{enumerate}
Note that the effect of the second accretion event leading to the recent rejuvenation on the disc may not be even across the disc. The inner disc is likely to be less affected than the outer disc since the lowest [Fe/H] in the low-$\alpha$ sequence increases toward the outer disc. 
A detailed study on the radial trend of the chemical abundances with high radial resolution will be presented in future work. 

In addition to the [Mg/Fe]--[Fe/H] plane, we also show comparisons of the distribution functions in [Fe/H] and [Mg/Fe] separately. 
Figure~\ref{hist-feh} shows the distribution in [Fe/H] (left-hand panel) and [Mg/Fe] (right-hand panel). The blue solid lines represent the observed distribution, while the orange dashed lines are the model predictions. 
The agreement is good, with a clear [Mg/Fe] bimodality in both model and observations.

To summarise, we propose a complex multi-phase scenario for the formation of the Milky Way's inner disc. In our model, the high-$\alpha$ sequence forms in an early starburst, while the metal-poor end of the low-$\alpha$ sequence originates from a second, more recent starburst. The metal-rich part of the low-$\alpha$ sequence, instead, forms gradually via a long-lived phase of low-level star formation between these two bursts. 


\subsubsection{Age}
\label{sec:datacomp}
In this section, we present a more detailed comparison between the model prediction and the observed data, including the information from stellar ages. Figure~\ref{agebin-model} shows the [Mg/Fe]--[Fe/H] plane for the old population ($t >8\;$Gyr) in the left-hand panel and the young population ($t<8\;$Gyr) in the right-hand panel. Grey and green contours show the observed and predicted distributions, respectively.

There is good agreement between model and observations for both age bins. It can be seen clearly that the high-$\alpha$ population is dominated by old stars, while the low-$\alpha$ populations is dominated by younger stars. On top of this general trend, the low-$\alpha$ sequence contains some old stars, leading to a bimodal distribution (left-hand panel). According to our simulation, this $\alpha$-dichotomy is intrinsic. The small fraction of young stars in the high-$\alpha$ population (right-hand panel), instead, is caused by the age uncertainty. A few intrinsically old stars with high-$\alpha$ abundance appear in the younger age bin due to the relatively large scatter in age.

\section{Discussion}
In this paper, we present a chemical evolution model for the evolution of the inner disc of the MW. The model is constrained by observational data from the APOGEE survey.

\subsection{Mg yields in the model}
\begin{figure*}
	\centering
	\includegraphics[width=16cm]{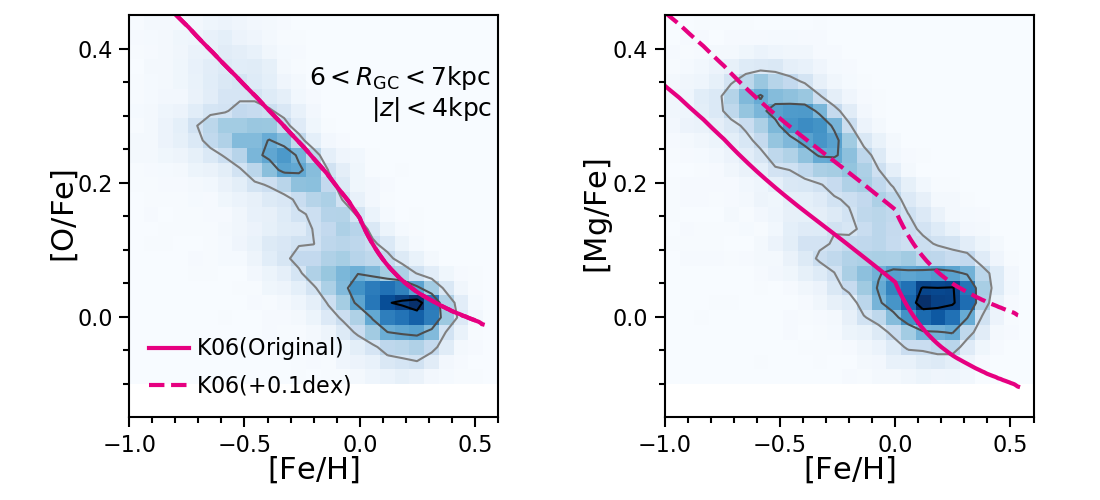}
	\caption{Comparison between model predictions (magenta lines) and observational data for the inner disc (density map and contours) in the [O/Fe]-[Fe/H] (left-hand panel) and [Mg/Fe]--[Fe/H] (right-hand panel) planes. The solid lines are models with the original SN-II yields from \citep{kobayashi2006}. The dashed line is a model in which the Mg-yield is enhanced by $0.1\;$dex. This Mg-enhanced model is used in the present paper.}
	\label{mg-test} 
\end{figure*}

By comparing the predicted ratio [$\alpha$/Fe] of different $\alpha$-elements to iron with observational data we notice that the magnesium production in the K06 SN-II yields may be underestimated. To illustrate this, Figure~\ref{mg-test} shows the comparison in the [O/Fe]-[Fe/H] and [Mg/Fe]--[Fe/H] planes using the original yields from \citet{kobayashi2006} (solid lines). The [Mg/Fe] ratio is considerably underestimated, while [O/Fe] fits the data well. 

This problem of SN-II Mg-yield calculation has already been highlighted by \citet{thomas1998}. We estimate the relative deficiency in Mg production by calculating the net offset between the model track and data in [Mg/Fe] with respect to [O/Fe] at a given [Fe/H], which is $-0.1\;$dex. To correct for this offset, in the present paper we increase the Mg-yield of \citet{kobayashi2006} by this amount. The dashed line in Fig.~\ref{mg-test} shows the model track with the enhanced Mg production. With this enhancement, the model is now able to match the observed oxygen and magnesium abundances simultaneously.

%

\subsection{The origin of the $\alpha$-dichotomy}
The chemical bimodality in the MW, found over two decades ago\citep{fuhrmann1998}, remains challenging, not only for phenomenological chemical evolution models \citep{spitoni2019}, but also cosmological simulations \citep{loebman2011,minchev2013}. 
A number of models in the literature only produce a broad continuous distribution in the [$\alpha$/Fe]-[Fe/H] plane (e.g., \citealt{loebman2011,minchev2013,naiman2018}).

Interestingly, in some recent simulations signatures of the $\alpha$-dichotomy begin to arise, {albeit with different origins}. {\citet{calura2009} presents the first chemical evolution model, embedded in a hierarchical semi-analytical model, that qualitatively reproduce the observed bimodal distribution in [$\alpha$/Fe]-[Fe/H] plane. The chemical bimodality arises due to a rapid transition between two star formation regimes, from an intense early star formation epoch to an inactive secular phase. In addition, delayed gas accretion also plays a role in forming the metal-poor branch of the low-$\alpha$ sequence. This multi-phase star formation history combined with delayed gas accretion is in good agreement with the disk formation scenario proposed in this work.}
\citet{grand2017} exhibits chemical bimodality in one of the six simulated galaxies, which is attributed to multiple star formation phases and disc contraction. \citet{mackereth2018} examine a sample of 133 Milky Way-like disc galaxies from the EAGLE simulation, and find that only $\sim 5$ per cent show a bimodal distribution in the [$\alpha$/Fe]-[Fe/H] plane. A common feature of these galaxies is that their gas accretion histories exhibit multiple episodes of accretion. \citet{clarke2019} present an interesting galaxy from a hydrodynamical simulation that shows chemical patterns at different spatial positions similar to that observed in the MW. An $\alpha$-bimodality is present in the inner disc owing to double-modes of star formation. Early, enhanced star formation produces the high-$\alpha$ stars, while an underlying low-level star formation activity forms the low-$\alpha$ population. 
{Similarly, in chemo-dynamical simulations by \citet{khoperskov2020}, a bimodal distribution in [$\alpha$/Fe]-[Fe/H] arises with the high- and low-$\alpha$ sequences formed during an early rapid phase and a subsequent secular phase of star formation, respectively.  
This dual-phase star formation history is qualitatively consistent with the SFH in this work without the rejuvenation stage.

Another possible origin of the $\alpha$-bimodality is a gas-rich merger that brings in metal-poor gas, dilutes the interstellar medium and stimulates star formation to form the low-$\alpha$ sequence \citep{buck2020}. This merger-induced formation of the metal-poor low-$\alpha$ sequence is supported by the age-abundance structure observed in the outer disc \citep{lian2020}. Alternatively, a major-merger event igniting star formation in a pre-existing metal-poor gaseous outer disc could also produce some level of bimodality in [$\alpha$/Fe]-[Fe/H] space \citep{agertz2020}.}
{While simulations are successful in reproducing the overall $\alpha$-bimodality, 
significant improvements are still needed to reach a good match with the observations in detail, such as the shape of the density and age distributions in the [$\alpha$/Fe]-[Fe/H] plane and their variations at different galaxy positions.} 

 
{The complex and time-consuming nature of hydrodynamical simulations limits its flexibility in simulating various galaxy formation histories and circumstances. The simple phenomenological model, as presented in this work, is very flexible and therefore serves as a complementary approach to interpret the detailed abundance observations in the Milky Way.} 
Based on a `two-burst' disc formation scenario, we successfully produce a bimodal distribution in [Mg/Fe]--[Fe/H] that matches the observations {well}. In this model, most stars in the high-$\alpha$ and low-$\alpha$ sequences originate from the two starbursts at early and late times, respectively. 
%
%
The key for the creation of the bimodality in our simulation is the rapid suppression of star formation following the first starburst. 
{Our results could be used as a guide for more complicated chemodynamical simulations to better reproduce the chemical pattern observed in the Milky Way and to probe the mechanisms driving the Milky Way's formation and evolution.}


\subsection{Thick disc formation}
Since the geometric thick disc of our Galaxy is largely dominated by the high-$\alpha$ sequence, thick disc formation must be closely related to the creation of the $\alpha$-dichotomy.

There are two main populations with distinct properties at the vertical distances of the thick disc ($1<z<4\;$kpc). One is the well-known old, high-$\alpha$ population ([$\alpha$/Fe]$>+0.2$) that dominates the inner part at radii $r<8\;$kpc. The other is a mildly $\alpha$-enhanced ([$\alpha$/Fe]$\sim+0.1$), young population, which is the metal-poor end of the low-$\alpha$ sequence. This population dominates at large radii, $r>10\;$kpc. The presence of these two populations at different radii of the thick disc leads to a radial gradient in many properties in the thick disc \citep{boeche2013,boeche2014,martig2016b}.

Interestingly, in our `two-burst' scenario, these two populations are exactly composed of the stars formed out of the two starbursts. This is not a coincidence. Stars formed during the starburst are likely born in a dynamically hot environment, which results in a large scale-height. The spatial separation of the two populations suggests that the early and late starbursts are spatially separated. In particular, the more recent burst preferentially affected the outer disc.   

What triggers the two starbursts? In our model, both starbursts are associated with the accretion of significant amount of gas on relatively short timescales. As discussed in Paper~I, the accretion of a gas-rich dwarf galaxy is a plausible explanation for the late accretion of gas, hence the second starburst.

The trigger of the first starburst is less clear. A gas-rich merger may be a possible scenario, but the merger event would need to be more violent because of the large scale-height of the high-$\alpha$ population in the inner thick disc (Bovy et al. 2012a, 2012b). Another possible trigger of the early starburst could be clumpy star formation caused by internal disc instability or galaxy interactions. Clumpy morphologies are found to be ubiquitous in high redshift galaxies in the early Universe (e.g., \citealt{elmegreen2005,ravindranath2006,guo2012}). \citet{clarke2019} produced a $\alpha$-dichotomy in a simulated galaxy involving clumpy star formation. 

The shutdown of gas supply and the rapid suppression of the SFE after the first starburst may well be the consequence of a dynamical cooling process. During this process, the thickened gas disc may fall back onto the mid-plane, and then fuel a low-level star formation activity to form a thin disc structure (the metal-rich low-$\alpha$ sequence of the inner thin disc today). This naturally explains the absence of young stars in the inner thick disc and the systematic shift of the high-$\alpha$ sequence at different vertical distances.

\subsection{Structure of mono-abundance populations}
The present model provides a natural explanation for the observed spatial structure of the mono-abundance population and the lack of structural bimodality.

Based on the analysis of the mono-abundance population in the [$\alpha$/Fe]-[Fe/H] plane, Bovy et al. (2012a, 2012b) derived the distribution of scale-height as a function of [$\alpha$/Fe] and [Fe/H]. {Surprisingly, scale-height} is found to decrease with decreasing [$\alpha$/Fe], with no hints of a bimodality (see also \citealt{mackereth2017}). According to our model, this decreasing scale-height has the same origin as the shift of the low- and high-$\alpha$ sequences in the [Mg/Fe]--[Fe/H] plane at various vertical distances.


The lack of structural bimodality is {likely} due to the {presence of metal-poor low-$\alpha$ stars which are formed during the} second, late-time starburst {at intermediate scale-heights}. 
{We speculate that excluding these stars 
could possibly help to recover {the discontinuity or even} bimodality in scale-height that coexists with the chemical bimodality.}



\subsection{Comparison with other models}
To explain observations from modern spectroscopic surveys, the original`two-infall' model by \citet{chiappini1997} has recently been modified \citep{grisoni2017,spitoni2019}. \citet{grisoni2017} propose a model in which the two phases of gas infall evolve independently and coevally in order to explain a group of $\alpha$-enhanced, metal-rich stars.  However, this model fails to produce the age distribution, especially for the low-$\alpha$ sequence, as shown in \citet{spitoni2019}. To solve this problem, \citet{spitoni2019} propose a further revision, in which the second phase of gas infall is delayed by $4.3\;$Gyr. This revised model produces a metallicity distribution that matches the observations overall. In addition the age distribution of the high-$\alpha$ population is now well reproduced. However, the model still falls short at predicting the correct age for the low-$\alpha$ population. 

A {more critical} challenge faced by these `two-infall' models is the bimodal distribution in $\alpha$-abundance and the significant lack of stars at intermediate [$\alpha$/Fe]. In these models, the SFR decreases gradually since the first major star formation phase. 
As a result, a significant fraction of stars with intermediate [$\alpha$/Fe] forms between the two major phases of star formation. The model in \citet{spitoni2019} produces too many of these stars, and therefore no double-peak in [$\alpha$/Fe] ({see} Fig.~8 in \citealt{spitoni2019}).  

One major improvement in our model compared to the `two-infall' models is that we adopt a more recent second-phase accretion, which gives rise to a younger low-$\alpha$ sequence that better matches the observed ages. As discussed in Paper~I, this recent episode of significant gas accretion can be interpreted as a gas-rich merger event, possibly with Sgr dwarf galaxy. Another critical element of the present model is the very rapid quenching of star formation after the first starburst. This process is essential to reproduce the bimodal distribution in the [$\alpha$/Fe]-[Fe/H] plane. 



\subsection{Room for improvement of the model}
Although the model presented here reproduces well the main observational features in the [Mg/Fe]--[Fe/H] plane, there is still room for improvement. As shown in Fig.~\ref{feh-afe-model}, the model predicts too many metal-poor stars (at [Fe/H]$\sim -1.0$). This suggests there may be a more rapid increase in SFR during the first starburst than that adopted in the model. 
Also, there is a density dip predicted by the model at [Fe/H]$\sim+0.1$ {in the low-$\alpha$ sequence} which is not observed. This feature is subject to change when assuming a different amount of gas accretion in the late-accretion event in the model. This modification is possibly needed to match the observations at different radii. 
Another interesting difference worth noting is the narrowness of the tracks predicted by the models, compared to the data, in both the low- and high-$\alpha$ branches. One possible reason for this discrepancy could be that the observational uncertainties adopted in the model simulation are underestimated. Another possibility is that the residual scatter is intrinsic, which suggests that the evolution path may be inhomogeneous in the disc within the radial range considered in this work. We aim to address these mismatches in future modelling, where we also consider radial dependence. 

Another possible reason for the mismatches described above may be the observational selection function, which could possibly bias the observed distribution. The APOGEE selection function estimated by \citet{bovy2016} is not yet considered in the current work. Further improvements along these lines, with a more quantitative determination of the model parameters and their uncertainties after considering the selection function, is the subject of future work. 

{As discussed in \textsection4.1, we do not consider galactic outflow or stellar radial migration in our model. The good match between our fiducial model and the observational data confirms that these two processes are not necessarily required to explain the main features in the abundance structure of the disc. However, we note that metal outflow could be important for our Galaxy at its early evolutionary stage. Thus including metal outflow would be an important improvement of our model when used to uncover the early evolutionary history of our Galaxy. }

\section{Summary}
We investigate the age-chemical abundance structure of the Galactic inner disc by studying the distribution of stars in [Fe/H] and [Mg/Fe], aided by stellar age information. 
The high- and low-$\alpha$ sequences slightly shift in the [Mg/Fe]--[Fe/H] plane at different heights, with systematically higher [Fe/H] and lower [Mg/Fe] in the plane. This is consistent with the trend of decreasing scale-length with decreasing [$\alpha$/Fe], as revealed by mono-abundance population analysis (Bovy et al. 2012a, 2012b).
By separating stars into old ($t>8\;$Gyr) and young ($t <8\;$Gyr) populations, we find that the bimodal distribution remains in the old population, but is much weaker in the young population. 
    
To explain the chemical bimodality, as well as to meet the age constraint, we propose a complex multi-phase inner disc formation scenario: two starbursts several Gyr apart. In our model an early starburst at a cosmic time of $\sim1.3\;$Gyr is rapidly quenched, leading to several Gyr of low-level star formation activity before the second burst at a cosmic times of $7\;$Gyr. The early starburst, possibly triggered by early merger events or disc instability, established the high-$\alpha$ sequence. The subsequent rapid suppression of star formation is likely associated with a dynamical cooling process of gas that falls back from a vertically extended structure onto the mid-plane. This rapid decrease in SFR leads to the shift in [Mg/Fe] from the high-$\alpha$ sequence. A small fraction of stars with intermediate [Mg/Fe] forms during this star formation quenching episode. 
This is the key to produce the $\alpha$-bimodality. 

In our model, the low-level star formation stage then lasts for $\sim4$\;Gyr, producing the metal-rich populations ([Fe/H]$>+0.1$) in the low-$\alpha$ sequence. At a cosmic time of $7\;$Gyr, the second starburst occurs. The stars that form during the second burst are iron-deficient due to the dilution of pristine gas accretion, and mildly $\alpha$-enhanced because of enhanced star formation during accretion. As a consequence, the second starburst produces the metal-poor population ([Fe/H]$<-0.1$) in the low-$\alpha$ sequence. The second accretion episode is a major gas accretion. In Paper~I, we discuss that this event may be caused by a minor merger with a gas-rich dwarf galaxy. 

Our model based on this multi-phase formation scenario reproduces the observed distribution in the [Mg/Fe]--[Fe/H] plane for the old and young populations remarkably well. 
%

Our two-starburst model provides a natural explanation for the formation of the geometric thick disc. Combining with the results from Paper I, we conclude that the two starbursts occurred at different radii. The first starburst formed inner disc stars within the solar radius, while the second starburst was more concentrated in the outer disc. 
Under the assumption that the stars formed during the two starbursts end up {with} large scale-heights, the first starburst produced the old, $\alpha$-rich, inner thick disc, while the second starburst produced the young, intermediate-$\alpha$ population in the outer thick disc. Radial gradients in [Fe/H], [$\alpha$/Fe], and age in the geometric thick disc naturally arise due to the different properties of stars formed during the two starbursts.
  
Our disc formation picture also provides an explanation for the structure of the mono-abundance populations. The stars formed during the secular formation phase between the two bursts characterised by low-level star formation activity will likely settle at small scale-heights, forming the thin disc. This naturally explains the observed trend of decreasing [$\alpha$/Fe] with decreasing scale-height, and the shift between the high- and low-$\alpha$ sequences in [Mg/Fe] and [Fe/H] at different vertical distances. The second starburst produces a large number of young stars with intermediate [$\alpha$/Fe] at large scale-heights, which reduces the significance of bimodality in chemical abundance but also in scale height.By excluding the metal-poor populations in the low-$\alpha$ sequence, a more significant structural bimodality is expected.

A thorough analysis of radial trends for both the thin and thick disc is the subject of future work.

\section*{Acknowledgements}
We are grateful to the referee for the useful report that helped improve the paper considerably. The Science, Technology and Facilities Council is acknowledged for support through the Consolidated Grant ‘Cosmology and Astrophysics at Portsmouth’, ST/N000668/1. Numerical computations were done on the Sciama High Performance Compute (HPC) cluster which is supported by the ICG, SEPnet and the University of Portsmouth. T.C.B.  acknowledges partial support from grant PHY 14-30152 (Physics Frontier Center / JINA-CEE), awarded by the U.S National Science Foundation.
J.G.F-T is supported by FONDECYT No. 3180210 and Becas Iberoam\'erica Investigador 2019, Banco Santander Chile. ARL acknowledges Chilean financial support for this research from FONDECYT Regular 1170476 and QUIMAL project 130001. DAGH and OZ acknowledge support from the State Research Agency (AEI) of the Spanish Ministry of Science, Innovation and Universities (MCIU)  
and the European Regional Development Fund (FEDER) under grant  
AYA2017-88254-P. R.R.M. acknowledges partial support from project BASAL AFB-$170002$ as well as FONDECYT project N$^{\circ}1170364$.

Funding for the Sloan Digital Sky Survey IV has been provided by the Alfred P. Sloan Foundation, the U.S. Department of Energy Office of Science, and the Participating Institutions. SDSS acknowledges support and resources from the Center for High-Performance Computing at the University of Utah. The SDSS web site is www.sdss.org.

SDSS is managed by the Astrophysical Research Consortium for the Participating Institutions of the SDSS Collaboration including the Brazilian Participation Group, the Carnegie Institution for Science, Carnegie Mellon University, the Chilean Participation Group, the French Participation Group, Harvard-Smithsonian Center for Astrophysics, Instituto de Astrofísica de Canarias, The Johns Hopkins University, Kavli Institute for the Physics and Mathematics of the Universe (IPMU) / University of Tokyo, the Korean Participation Group, Lawrence Berkeley National Laboratory, Leibniz Institut für Astrophysik Potsdam (AIP), Max-Planck-Institut für Astronomie (MPIA Heidelberg), Max-Planck-Institut für Astrophysik (MPA Garching), Max-Planck-Institut für Extraterrestrische Physik (MPE), National Astronomical Observatories of China, New Mexico State University, New York University, University of Notre Dame, Observatório Nacional / MCTI, The Ohio State University, Pennsylvania State University, Shanghai Astronomical Observatory, United Kingdom Participation Group, Universidad Nacional Autónoma de México, University of Arizona, University of Colorado Boulder, University of Oxford, University of Portsmouth, University of Utah, University of Virginia, University of Washington, University of Wisconsin, Vanderbilt University, and Yale University.

\section*{Data Availability}
The data underlying this article is from SDSS-IV/APOGEE Data Release 16, which are available at \url{https://data.sdss.org/sas/dr16/apogee/spectro/aspcap/r12/l33/allStar-r12-l33.fits}, with datamodel at \url{ https://data.sdss.org/datamodel/files/APOGEE_ASPCAP/APRED_VERS/ASPCAP_VERS/allStar.html}. The chemical evolution model results are available upon request.

\end{document}